\documentclass[preprint,pre,a4paper,superscriptaddress,showpacs,nofootinbib]{revtex4}

\usepackage{graphicx}
\usepackage[body={16cm,22cm},bottom=1.5cm]{geometry}
\usepackage{amsmath}
\usepackage{latexsym}
\usepackage{amsfonts}
\usepackage{psfrag}
\usepackage{epstopdf}

\begin{document}
\title{Local dynamics and primitive path analysis for a model polymer melt near a surface}
\author{Mihail Vladkov\footnote{To whom correspondence
should be addressed. E-mail: mihail.vladkov@lpmcn.univ-lyon1.fr},
Jean-Louis Barrat}

\affiliation{  Laboratoire de Physique de la Mati\'ere Condens\'ee
et Nanostructures
               Universit\'e Lyon 1; CNRS; UMR 5586
               Domaine Scientifique de la Doua
               F-69622 Villeurbanne cedex; France
}

\date{\today}
\setcounter{page}{1}

\begin{abstract}
Using molecular dynamics simulations, we apply primitive path and
local Rouse modes analysis to study the chains conformations and
the local dynamics and viscosity of a model polymer melt near a
flat repulsive wall and a repulsive wall presenting some bonding
sites. The presence of a repulsive wall leads to acceleration of
the dynamics both for unentangled and weakly entangled melts and
to a depletion in the entanglement density in the wall vicinity.
When the surface bears grafted chains, we show that the melt
chains are accelerated in the unentangled regime and slowed down
in the entangled regime. By analyzing the primitive paths we
show that the observed slowdown in presence of grafted
chains is due to an increase in the entanglement density in the
interfacial layer. In contrast, for a plain repulsive surface 
we observe a depletion of the entanglements at the interface.
The presence of a relatively small density of
grafting sites thus leads to improved mechanical properties
(reinforcement) and decreases locally the entanglement length even
if the surface is repulsive.

\end{abstract}

\maketitle

\newpage
\section{Introduction}
Nanocomposite polymer based materials have many interesting
properties leading to their extensive usage in practical
applications. It has been long known that adding filler particles
to a polymer melt can lead to substantial modifications in its
mechanical properties. The properties of polymer-filler composites
are often associated with filler clustering and percolation
\cite{payne,oberdisse} but several  studies over the past several
years showed that the effect is also observable below the
percolation threshold \cite{sternstein,gautier}. This indicates
that the particle-matrix interface is another crucial ingredient
in the complex physics of filled elastomers. Another indication of
the importance of surface effects lies in the fact that small size
particles (nanoparticles) have a much greater effect compared to
micron size fillers.  The nature of the polymer filler
interactions was shown to be very important, it was established by
experiment \cite{lequeux} and simulation \cite{bashnagel-review}
that attractive interactions lead to an increase of the glass
transition temperature near the interface. With this argument
reinforcement can be partly explained by formation of ``harder'',
glassy layers with slow dynamics around attractive fillers. Most of the
fillers used in practical applications (tires) and leading to
substantial reinforcement, such as treated silica particles,
exhibit a globally repulsive interface presenting some sites that
can covalently bond polymer chains from the melt. The physics of
reinforcement on the interface in this case remain unclear at a
microscopic scale. Experimental studies \cite{sternstein}
suggested that chain entanglements in the surface vicinity should
play a major role in the reinforcement and it can be explained by
the presence of trapped entanglements near the interface.
Theoretical studies in the past qualitatively suggested that the
density of entanglements should decrease in the vicinity of a wall
\cite{brown1,brown2} because of the average decrease in chain
dimensions established earlier by simulation
\cite{kumar-Rg,binder-Rg}. On the experimental side the validity
of this assumption was recently questioned for free standing films
\cite{Itagaki}, but direct evidence of this phenomenon near a
repulsive wall is still lacking, and it could not be related to
mechanical properties.

Recent advances in molecular simulation has made possible a
precise study of these problems. The well established concept of
entanglement in a polymer melt could be ``observed'' by direct
primitive path analysis of a polymer melt
\cite{everaers,sukumaran}. We recently developed an equilibrium
method for studying the mechanical properties of a melt that can
be used to assess local viscoelastic properties
\cite{vladkov-barrat}. Using these methods we analyze the dynamics
and the entanglement density of a polymer melt near a repulsive
wall with or without bonding sites and give a generic explanation
of the melt slowdown or acceleration at the interface as a
function of  molecular weight.

The systems under study are briefly described in the next section.
We then discuss the local dynamics of the melt in terms of the
Rouse modes of the chains for unentangled and weakly entangled
melts. Finally, we   study the structure of primitive paths in the
vicinity of the wall, in order to explain the observed dynamical
behavior.

\section{System description and methods}
\subsection{Model}
The chains are modelled by an abstract and generic, though well
studied, bead spring model - the rather common "Lennard-Jones +
FENE" model\cite{KremerGrest}. All monomers in the system are
interacting through  the Lennard-Jones potential:
\begin{equation}
\label{eq:ljpotcut}
U_{lj}(r) =  \bigg\{ \begin{array}{lll}
         4\varepsilon((\sigma/r)^{12} - (\sigma/r)^{6}), &r\le r_c \\
         0, &r>r_c
       \end{array}
\end{equation}
where $r_c=2.5\sigma$. Neighboring monomers in the same chain are
linked by the FENE (Finite extension non-linear elastic)
potential:
\begin{equation}
\label{eq:FENE}
U_{FENE}(r) =  \frac{k}{2}R_0 \ln(1-(\frac{r}{R_0})^2), \qquad r<R_0
\end{equation}
where $R_0=1.5\sigma$ and $k=30.0 \varepsilon/\sigma^{2}$.

The temperature of the melt was fixed in all simulations at $k_B T
= 1$, well above the glass transition temperature for our model.
We studied systems of chain length of 10, 20, 50 and 100 beads
with a total number of beads ranging from 6400 to 51200. This
allows us to investigate the crossover into the weakly entangled
regime, with an entanglement mass $N_e$ estimated for this model
in the vicinity of 50\cite{everaers}. The melt was confined
between walls in the $z$ direction with $L_z
> 5R_g$ in all cases. The interaction between the wall and the
beads was chosen to crudely reproduce the PE - silica interaction.
The bead diameter was mapped through the polymer $C_{\infty}$
ratio \cite{Flory} to give $\sigma \sim 8 \AA $. We apply a mixing
rule using values for the PE and silica interaction intensity
found in the literature \cite{mapping} to obtain a Lennard-Jones
potential between the polymer and the wall of $\sigma_{wall} =
0.6875\sigma$ and $\varepsilon_{wall} = 0.82\varepsilon$. The
potential is cut off at its minimum $2^{1/6}\sigma_{wall}$ so that
it is purely repulsive, as expected for the interaction between PE
and an untreated silica surface. The wall is represented either by
a flat potential or by the  111 surface of an FCC lattice made of
spherical particles, supplemented by a flat repulsive potential in
the second layer, that  prevents beads from escaping. For each
chain length two systems were prepared: one with a purely
repulsive wall and one with chemisorbing sites on the wall, where
grafted chains are anchored. Silica surface treatment consists in
introducing very short chain molecules on the surface that
covalently bond with some sites on the surface on one side and
with monomers on the other side. The number of bonding sites in
the simulation was calculated assuming that $4\%$ of the silica
molecules covering the surface of the wall are active and react
with the bonding molecule.\cite{mapping} The surface density of
active sites is therefore  $0.2\sigma^{-2}$. The resulting bond
between a wall particle and a coarse grained bead is thus a soft
entropic spring of finite length of the order of several chemical
units that is modelled by an non-harmonic spring:
\begin{equation}
\label{eq:bond}
U_{bond} = \frac{\varepsilon_{bond}(r-r_0)^2}{[\lambda^2-(r-r_0)^2]}
\end{equation}
The spring constant was set to $\varepsilon_{bond} = \frac{3}{2}k_BT = 1.5$,
the equilibrium distance and the finite extension length are set
to $r_0 = \lambda = 0.8\sigma$. These parameters define a soft spring
freely fluctuating with ambient temperature that cannot extend further
than $1.6\sigma$ away from the surface. All the simulations were performed
using the LAMMPS code \cite{lammps}.

\subsection{System Preparation and Static Properties}

\begin{figure}[ht]
\centering
\includegraphics[width=7cm]{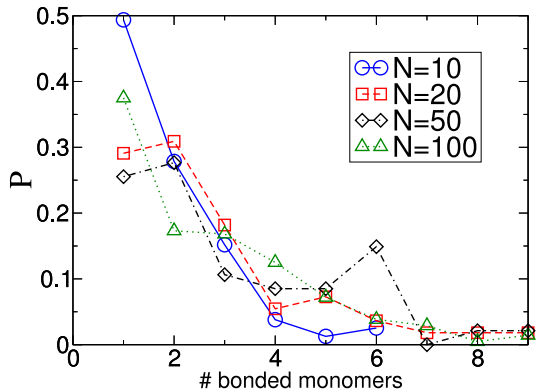}
\caption[Distribution of the number of bonded monomers]
{Distribution of the number of bonded monomers per grafted
chain in the systems with treated surface}
\label{fig:p_na}
\end{figure}

\begin{figure}[ht]
\centering
\includegraphics[width=7cm]{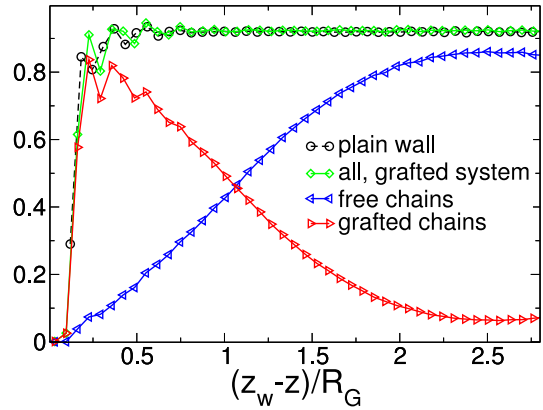}
\caption[Density profile for N=100]
{Monomer density profile of the system with and without grafted chains for N=100.
The densities of the monomers of the grafted and the free chains are also shown.
The density profiles in the other systems have similar behavior.}
\label{fig:dens}
\end{figure}

At first a pure polymer melt was confined at a given pressure
(around $0-0.8$ depending on the system) between two repulsive
walls. Pressure is monitored by calculating the normal force on the walls.
Then, in order to obtain the system with grafted chains from this
initial configuration, a new bond is created between each active wall site and
the closest monomer of the melt. Only one bond per active wall site is allowed
while a monomer can be bonded to several wall atoms as it is a
coarse-grained bead representing several chemical units ($1bead \sim 6 CH_2$).
There is no preferential bonding for end or middle monomers.
The new system is now equilibrated at the same pressure as the original
system, its pressure being calculated via the force on the wall and the grafted chains.
All systems were equilibrated for $10^6$ ($N=10$) to $10^7$ ($N=100$) time steps before
production runs. The grafting procedure creates a population of grafted chains near
the wall having around $3.5$ grafted beads per chain (fig. \ref{fig:p_na}).
The grafted chains extend in all systems a distance of maximum $2.5R_G$ in the
melt (fig. \ref{fig:dens}). There are no bridges between the walls.

\begin{figure}[ht]
\centering
\includegraphics[width=7cm]{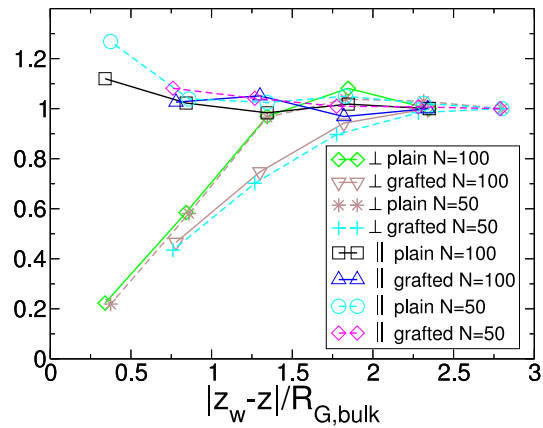}
\caption[ radius of gyration N=50 and N=100] {Variation of the
radius of gyration squared, as a function of the distance to the
wall divided by $R_G$. The results  are normalized by the bulk
value $\langle R_G^2 \rangle$} \label{fig:RGz}
\end{figure}

We start with a brief study the influence of the surface in the
two configurations on the static properties of the chains. The $z$
component of the mean end to end vector and the radius of gyration
of the chains decreases in the vicinity of the wall and there is a
very slight increase in the size  of the chains parallel to the
wall (see fig. \ref{fig:RGz}). The effects are observed to extent
within  about one  $R_G$ (bulk value) from the wall, as was
established in earlier simulations. \cite{binder-Rg,kumar-Rg}

In  systems with grafted chains the decrease extends farther into
the melt, up to $2R_G$ from the wall surface. This slightly larger
length scale is close to the typical spatial extension of the
grafted chains, that are on average more extended than the bulk
chains.

\section{Local dynamics and viscosity of polymer chains}\label{dynamics}

\subsection{Desorption and Mean Square Displacement}
\begin{figure}[ht]
\centering
\includegraphics[width=7cm]{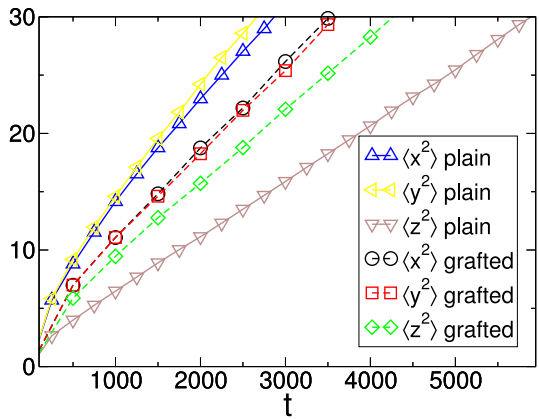}
\caption[Mean square displacement of monomers near the wall]
{Monomer mean square displacement in the wall vicinity for the
system $N=20$ with and without grafted chains.}
\label{fig:msd}
\end{figure}

We can get a qualitative idea of the melt dynamics in the bulk and
near the surface by looking at the local mean square displacement of
the polymer chains. We define the bulk and surface chains mean square displacement
by:
\begin{eqnarray}
\langle R^2(\tau) \rangle_{bulk} &=& \frac{1}{TN_{ch}^z}\int dt \sum_{i;\; -1 < Z_{cm}^i(t) < 1}(R_i(t+\tau)-R_i(t))^2 \\
\langle R^2(\tau) \rangle_{wall} &=& \frac{1}{TN_{ch}^z}\int dt \sum_{i;\; z_w-2 < |Z_{cm}^i(t)| < z_w}(R_i(t+\tau)-R_i(t))^2
\end{eqnarray}
In the case of a plain surface we find that surface chains have
increased mobility parallel to the surface in the $x$ and $y$
direction compared to bulk chains and they are slowed down
perpendicular to the surface. For a wall with grafted chains the
free chains in the wall vicinity are slower in the parallel
direction (fig. \ref{fig:msd}) - this is attributed to the
presence of the grafted chains acting as obstacles for  motion
parallel  to the wall.

A more surprising result is that, in the presence of grafted
chains,  the free chains are faster than 
the chains near a plain wall in the
perpendicular direction (fig. \ref{fig:msd}). This kind of
behavior can be understood knowing the desorption mechanism from a
flat surface \cite{kurtsmith}. We know that, even in the case of a
flat repulsive wall, surface chains having relaxed conformations
with many monomers on the surface 
are very slow to desorb and are thus responsible for a slow
down in dynamics near the surface. This is seen in the
perpendicular slowdown in the plain wall system. In the presence
of grafted chains, there are less fully relaxed free chains near
the surface, as the corresponding conformations are achieved by
the grafted chains that do not desorb. The presence of grafted
chains makes the surface rough on a scale comparable to the
polymer size, it can be argued that adsorption and desorption on a
rough surface for polymers is faster as it involves less entropy
loss than in the case of a flat wall \cite{huber}. Thus the free
chains population near the surface has faster exchange dynamics
with the bulk chains.

\subsection{Rouse Modes}
We study the local dynamics of the polymer chains by monitoring the
relaxation of the Rouse modes of the chains:
\begin{equation}
\label{eqn:rmodes}
X_p(t) = \frac{1}{N}\sum_{n=1}^N r_n (t) \cos{\left( \frac{(n-1/2)p\pi}{N}\right)}\,,\;\;p=0,\ldots,N-1
\end{equation}
with $N$ the chain length and $r_n(t)$ - the position of the n-th
monomer in the chain at the time $t$. The correlation of the
$p$-th Rouse mode describes the relaxation of a subchain of $N/p$
monomers, so that the study of this single chain quantity allows
to probe the dynamics on different length scales. Being a local
single chain quantity, the correlation of the Rouse modes allows
us to investigate the dynamics in different sub-volumes of a
non-homogeneous system, provided  they contain a large enough
number of chains. It is also a route to calculate the local
viscosity in each region, as discussed in ref.
\cite{vladkov-barrat}. In our confined systems, we define the
local modes relaxation as a function of the $z$ coordinate as:
\begin{eqnarray}
\langle X(\tau)X(0) \rangle(z) = \frac{1}{TN_{ch}^z}\int dt \sum_{i;\; z-dz < Z_{cm}^i(t) < z+dz}X_i(t+\tau)X_i(t)
\end{eqnarray}
It is expected that the slowest Rouse modes relaxes  on a time
scale smaller than the time needed for a chain to diffuse its own
size. Hence, we choose a slice width of the order of $R_G$, so
that the center of mass of a chain stays (statistically) within
the same slice while the correlation is measured. This allows us
to access local dynamics with a spatial resolution of the order of
$R_G$, which is better than the resolution associated with
measurements of the  mean square displacement. For the systems
with grafted chains we define relaxation times by considering {\em
only the modes of the free chains}. Hence the values of the
relaxation times shown in this study are not affected directly by
the frozen dynamics of the tethered chains. Our goal is to
understand the influence of the presence of the grafted chains on
the remaining melt and to capture, if any, matrix mediated slow
down.

\begin{figure}[ht]
\centering
\includegraphics[width=7cm]{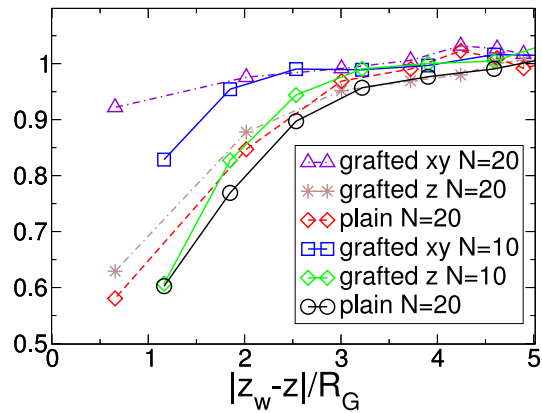}
\caption[Normalized Rouse times N=10 and N=20]
{Local Rouse times normalized by the bulk value
for chain lengths of 10 and 20 and for the free chains in the systems
with grafted chains (referred to as ``grafted'') 
and for the chains in the system without grafted chains
(referred to as ``plain'').}
\label{fig:norm-rtimesN1020}
\end{figure}

We estimate the Rouse times by an exponential fit of the
normalized correlation function of the first mode, which have in
all cases a clear exponential behavior. We estimate separately the
relaxation of chain conformations following the three directions
in order to take into account the spatial inhomogeneity near the
surface. For chain lengths of $N=10$ and $N=20$ we observe a clear
decrease in the relaxation times of the modes near the wall in all
systems, regardless of the presence of grafted chains (fig.
\ref{fig:norm-rtimesN1020}). In the case of a flat repulsive
surface this acceleration is expected and can be understood in
terms of reduction of the monomeric friction due to the repulsion
of the surface \cite{bashnagel-review}. We also studied the
influence of the pressure and the microscopic wall roughness on
the acceleration of the chains. They were found to have a
negligible effect on the modes relaxation in the melt. Cage
effects associated with microscopic roughness of the wall are
relatively unimportant for chain molecules compared to simple
fluids, due to the competition with bond constraints associated
with the neighboring monomers in the chain.

As the grafted chains act as soft obstacles for the free chains
one could expect that the free chains will be slower close to the
wall than in the bulk. Studying the local Rouse times we find that
this assumption is not true for  melts with chain lengths
$N=10,20$. In the systems where there are attached chains on the
surface, the free chains in the immediate vicinity of the wall are
accelerated with respect to the bulk chains. The effect of the
grafted chains is seen when comparing the systems with and without
grafted chains.
In the direction parallel to the surface the free chains in the
immediate vicinity of the wall having grafted chains are about
$20\%$ ($N=10$) and $30\%$ ($N=20$) slower compared to the chains
seeing a plain repulsive surface, but they are still accelerated
by $20\%$ ($N=10$) and $10\%$ ($N=20$) compared to the bulk
dynamics (see fig. \ref{fig:norm-rtimesN1020}). In the
perpendicular direction the chain conformations equilibrate in
about the same time for the two types of systems. For these
systems of short chains, the slowing down due to the grafted
chains is not sufficient to overcome the acceleration due to the
repulsive surface and we do not expect reinforcement with respect
to the bulk properties of the material, as discussed in the next
section.

\begin{figure}[ht]
\centering
\includegraphics[width=7cm]{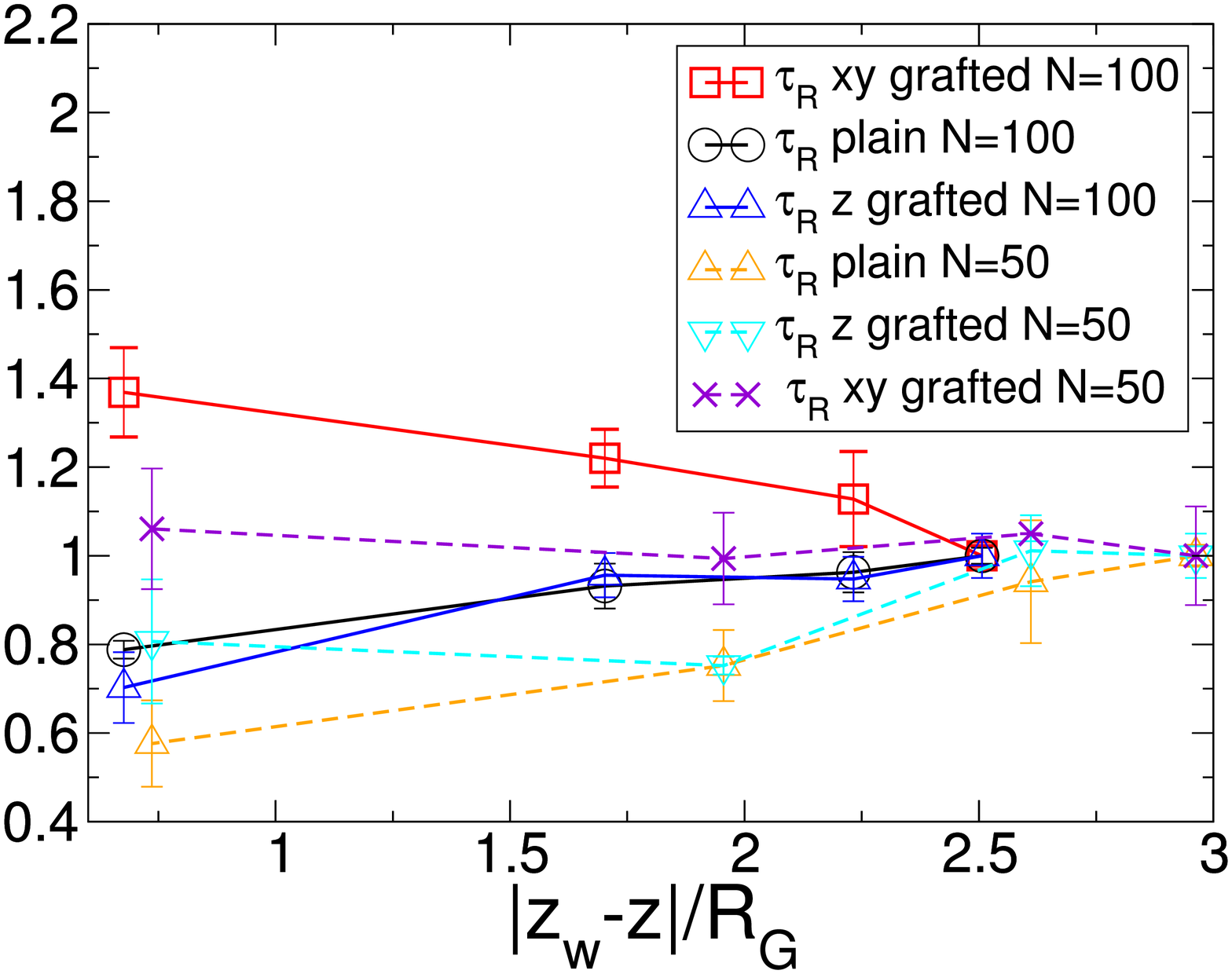}
\caption[Normalized Rouse times N=50 and N=100] {Local Rouse times
(first mode), normalized by the bulk value, for chain lengths of
50 and 100 and systems with and without grafted chains.}
\label{fig:norm-rtimesN50100}
\end{figure}

\begin{figure}[ht]
\centering
\includegraphics[width=7cm]{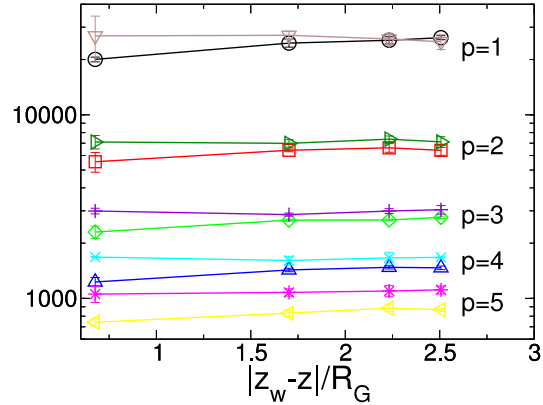}
\caption[Relaxation times of the first 5 Rouse modes, N=100]
{Local relaxation times of the first five Rouse modes for chain
length of 100 for the systems with  and without grafted chains.
The system without grafted chains is systematically the faster
one.} \label{fig:5modes_increase}
\end{figure}

For chain lengths $N>20$ we measure the modes relaxation times as
the integral of the normalized correlation function so that an
exponential behavior of the latter is not required. Although the
relaxation is not strictly exponential, the resulting values are
close to those that would be obtained using an exponential fit.
For chain lengths around and above the entanglement length we
observe an acceleration for the first Rouse mode in the case of a
plain wall, very similar to the case of short chains.  The
influence of grafted chains, however, is much more important than
for short chains.  In the systems with grafted chains of
$N=50,100$, we measure a slowing down for the largest relaxation
time in the parallel direction (of around $10\%$ for $N=50$ and
$40\%$ for $N=100$ within a distance of $R_G$ from the wall) with
respect to the dynamics in the middle of the film (see fig.
\ref{fig:norm-rtimesN50100}). The average relaxation times in the
layer feeling the presence of the grafted chains are larger than
those for the system without grafted chains. On average, for
$N=100$, the mean relaxation times of the first five Rouse modes
are increased by $\sim 20\%$ in the presence of grafted chains,
compared to what is observed for  a plain wall (see fig.
\ref{fig:5modes_increase}). In summary, the presence of grafted
chains induces a slowing down for all modes (compared to the plain
wall case), and in the case of the first mode a slowing down
compared to the bulk is  also observed.

In the $z$ direction the conformation dynamics varies in a similar
way as a function of the distance to the surface, regardless of
the presence of grafted chains.

\begin{figure}[ht]
\centering
\includegraphics[width=7cm]{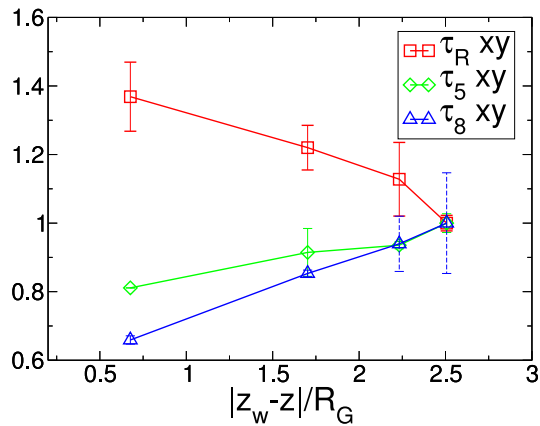}
\caption[Modes relaxation times N=100 with grafted chains]
{Local relaxation times (N=100) normalized by the bulk value of the first (entangled) mode and of higher modes
for the free chains in the system with grafted chains.}
\label{fig:highp}
\end{figure}

\begin{figure}[ht]
\centering
\includegraphics[width=7cm]{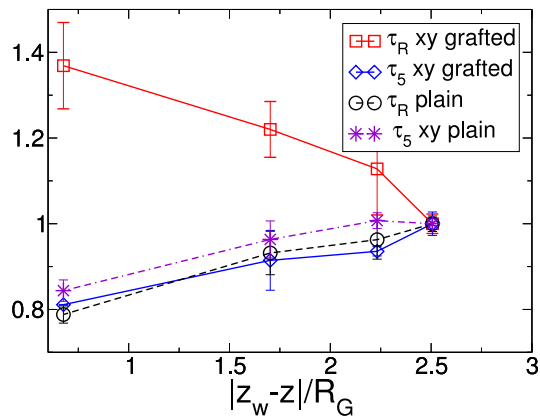}
\caption[Modes relaxation times N=100 with grafted chains and plain surface]
{Relative variation of local relaxation times normalized (N=100) by the bulk value of the first (entangled) mode and the fifth mode
for the free chains in the system with grafted chains in comparison with the
local Rouse times for a plain flat surface}
\label{fig:highp2}
\end{figure}

In the following, we argue that the substantial  slowing down
observed in the relaxation time of the first mode  is due to
entanglement effects in the surface vicinity.  If the  slowing
down is due to entanglements, we should still have faster dynamics
close to the surface for chain segments well below the
entanglement threshold. In other words, the relaxation of a mode
$p$ such that $N/p > N_e$ should be slowed down when approaching
the wall, while the dynamics of a mode $p$ satisfying $N/p < N_e$
should be similar to the unentangled case, i.e. accelerated with
respect to its value in the middle if the film. This assumption is
confirmed by the measurement of the relaxation of higher Rouse
modes of the chains (see fig. \ref{fig:highp} and fig.
\ref{fig:highp2}). In terms of effective monomeric friction the
presence of the repulsive surface still leads to relative
acceleration of the dynamics in the immediate vicinity of the wall
on length scales smaller than the tube diameter (as in the case of
unentangled chains). This leads to a relative acceleration  of the
small scale modes in the interface layer when approaching the
wall. On the scale of entanglements the shorter the distance to
the wall is, the slower are the dynamics. This issue will be
further investigated in terms of entanglements in the next
section.

In summary, the presence of a repulsive surface results in the
formation of an interfacial layer where the dynamics and chain
conformations are different from the bulk. Chains tend to lie flat
on the surface so that their dimension perpendicular to the
surface is reduced in a layer extending one to two gyration radii
from the surface. The presence of the surface also alters the
dynamical properties of the melt in its vicinity within around two
to three bulk $R_G$. The presence of grafted chains on the surface
increases the thickness of the interfacial layer as far as static
properties are concerned due to the tendency of the grafted chains
to have an extension larger than the bulk $R_G$.

In the unentangled regime, the presence of a repulsive surface
results in a relative acceleration of the melt dynamics when
approaching the wall regardless of the presence of grafted chains
on the surface. The presence of grafted chains (below the polymer
brush density) slows down the free chains dynamics parallel to the
surface compared to the system without grafted chains but chains
are  accelerated with respect to the bulk.

The situation changes when the chain length increases and we enter
the entangled regime. In the case of entangled melts, the
relaxation of entangled modes $p$ such that $N/p > N_e$ is slowed
down compared to the bulk near the surface in presence of grafted
chains. In the next section we will relate this to the local
entanglement density. The local dynamics at length scales shorter
than the entanglement length is slower with grafted chains than
without, but remains accelerated close to the surface with respect
to the bulk.

In the perpendicular direction adsorption-desorption dynamics are
slightly accelerated compared to the system without grafted
chains. This is due to the fact that the slowest chains in the
vicinity of a flat  wall tend to be those with most contacts with
the wall \cite{kurtsmith}. In the grafting process, these chains
are removed from the population of "free" chains that we consider
here.

\subsection{Local Viscosity}

\begin{figure}[htbp] 
   \centering
   \includegraphics[width=7cm]{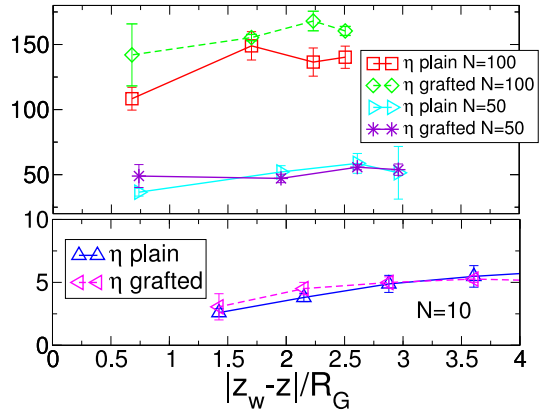}
   \caption{Local viscosity for the different systems. The
values are calculated as the mean of the
three components $xy$, $xz$ and $yz$ of the viscosity and
error bars indicate the dispersion around the mean value.}
   \label{fig:avg_visc}
\end{figure}

Next we measure the local viscosity in our systems using the method
discussed in ref. \cite{vladkov-barrat}. Each mode $p$ of the chains
has a contribution to the stress auto correlation function:
\begin{equation}
G_p (t) = \frac{\rho k_B T}{N} \frac{\langle X_{p\alpha}(t) X_{p\beta}(t) X_{p\alpha}(0) X_{p\beta}(0) \rangle}{\left(\frac{\langle X_{p\alpha}^2 \rangle + \langle X_{p\beta}^2 \rangle}{2}\right)^2}
\end{equation}
where $\alpha , \beta = x,y,z$. The mode contribution to the
viscosity of the mode is $\eta_p= \int_0^{\infty}G_p (t)dt$ and
the total viscosity is given by $\eta = \eta_{p=1}+ ... +
\eta_{p=N-1}$. For chain lengths of $N=10$ and $N=20$ the
viscosity is estimated from all the Rouse modes and for chains of
$N=50$ and $N=100$ only the first ten modes were considered. The
contribution of the eleventh mode for $N \ge 50$ was found to be
less than $1\%$ of the contribution of the first mode. The
contribution of the non polymeric stress  (associated with short
range interactions between monomers rather than with chain
connectivity) on very short time scales was neglected. While this
leads to an underestimate of the viscosity of the order of $20\%$
in the unentangled regime,  this contribution is negligible
($<1\%$) for longer chains as discussed in \cite{vladkov-barrat}.
The average local viscosity is shown in fig. \ref{fig:avg_visc}.
For chains of length $N=100$, the grafting induces an increase in
the local   viscosity in a range of around $2R_G$ from the wall,
as could be expected from the increase in the relaxation times. A
smaller increase in a layer of $R_G$ is observed for the chains
around the entanglement threshold ($N=50$). For unentangled melts,
there is essentially no difference between the results with and
without grafted chains.

\section{Primitive Path Analysis}
\begin{figure}[htbp] 
   \centering
   \includegraphics[width=7cm]{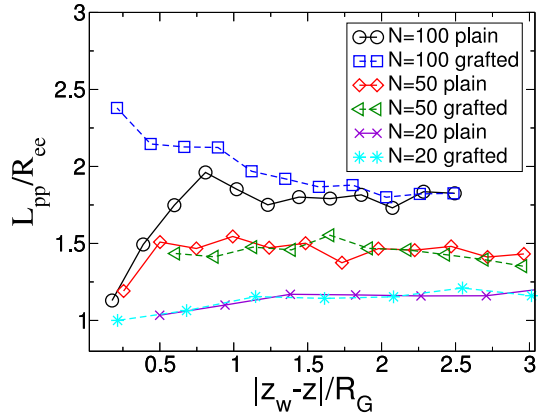}
   \caption{Local ratio of the primitive path length to the end-to-end
chain distance in the different systems.}
   \label{fig:lpp-ree}
\end{figure}

In order to interpret  the  slowing down in the dynamics for
entangled systems in the presence of grafted chains, we perform a
local primitive path analysis following the algorithm discussed in
ref. \cite{sukumaran}. Starting from independent initial states
separated by more than a chain relaxation time, the chain ends are
kept fixed, while the intra chain pair interactions are switched
off and the bond length is reduced to zero while increasing the
bond tension to $k=100$. In the systems with grafted chains the
grafting bonds are maintained, the primitive path quench is
applied to all the chains and the parameters of the primitive
paths are measured for the free chains only as in the previous
section. We measure locally, as a function of the distance to the
surface, the length of the primitive paths ($L_{pp}$). If no
entanglements exist between the chains the length of their
primitive paths should be equal or very close to their end-to-end
distance. The presence of entanglements leads to primitive paths
longer than the end-to-end distance with a typical Kuhn length
$a_{pp} = \langle R^2 \rangle / L_{pp}$ and an average bond length
$b_{pp} = L_{pp}/N$, which are related to the entanglement length
\cite{everaers}. Thus the entanglement length is given by:
\begin{equation}
N_e=\frac{a_{pp}}{b_{pp}} = \frac{N \langle R^2 \rangle}{L_{pp}^2}
\label{eqn:ne}
\end{equation}
and the average number of entanglements per chain is
$\frac{L_{pp}}{a_{pp}} - 1$. For chain lengths of 10 and 20 the
primitive path analysis shows that we are in the unentangled
regime as expected (with $0.1$ entanglements per chain for
$N=20$). For $N=50$ the melt is closer but still under the
entanglement threshold with $0.5$ entanglements per chain. For a
chain length of 100 there are already $1.1$ entanglements per
chain and the system is weakly entangled. As expected in the
unentangled case (fig. \ref{fig:lpp-ree}), the behavior in terms
of primitive paths of the free chains in the system with and
without grafted chains is identical. There is a slight decrease of
the primitive path length close to the surface that is related to
the decrease in the dimensions of chains lying flat on the
surface. A noticeable difference is seen in the entangled case
where the ratio $L_{pp}/R_{ee}$ becomes larger for the free chains
near the surface bearing tethered chains, while it decreases with
respect to the bulk value for the plain wall system.

\begin{figure}[htbp] 
   \centering
   \includegraphics[width=7cm]{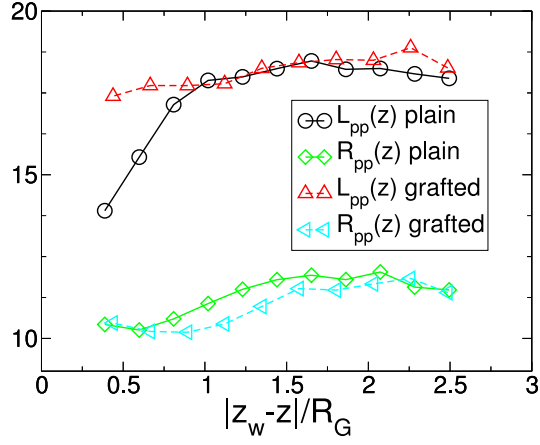}
   \caption{Local primitive path length and end-to-end
distance for $N=100$.}
   \label{fig:l-and-r}
\end{figure}

Measuring separately the primitive path length and the end-to-end
distance in the systems of $N=100$ (see fig. \ref{fig:l-and-r}) we
see that in the presence of grafted chains on the surface the
length of the primitive paths of the free chains remains
essentially constant throughout the boundary layer, while it is
decreased in the presence of a plain wall. A the same time there
is a slightly more pronounced decrease in the end-to-end distance
for the system with grafted chains as discussed previously (fig.
\ref{fig:RGz}, \ref{fig:l-and-r}). The grafted chains, being on
average more extended than the free chains provide more
entanglements in the plane parallel to the wall for the chains
close to the interface, thus keeping their path length constant
regardless of the fact that the perpendicular size of the chains
diminishes leading to less interpenetration in the $z$ direction.
In the case of a plain surface the decrease in the chains $R_G$
results in less interpenetration diminishing the primitive paths.

\begin{figure}[htbp] 
   \centering
   \includegraphics[width=6cm]{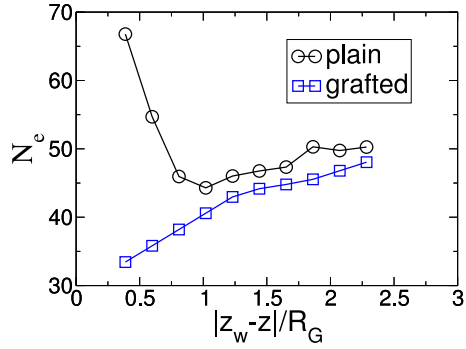}
   \caption{Local entanglement length for the two systems
of $N=100$. $N_e (z)$ was calculated from the simulation data
using equation \ref{eqn:ne}}
   \label{fig:ne}
\end{figure}

These effects can be interpreted in terms of local reduction of
the entanglement length (or, equivalently increase in the
entanglement density) as shown in fig. \ref{fig:ne}. The bulk
entanglement length found (around $N_e=50$) is in qualitative
agreement with the result of reference \cite{everaers}. In a
region that extends about one  $R_G$ from the surface, there is a
depletion of entanglements for a plain repulsive wall, as
qualitatively predicted in reference \cite{brown1}. This effect
was associated with smaller chain size in the $z$ direction. It
can be also understood knowing that chains in the immediate
vicinity of the wall only have neighboring chains on one side and
no chains to entangle with on the other side, so they have a
smaller total number of entanglements. The difference in
entanglement density explains the slowing down in the dynamics of
the entangled modes. Moreover, a quantitative prediction can be
established. Knowing from reptation theory that the relaxation of
an entangled chain is proportional to $N\times
\left(\frac{N}{N_e}\right)^2$, the ratio between the relaxation
times of two systems of equal chain length should be equal to the
inverse square ratio of the entanglement lengths in the two cases,
i.e.
\begin{equation}
\frac{\tau_R^{grafted}}{\tau_R^{plain}} = \left(\frac{N_e^{plain}}{N_e^{grafted}}\right)^2
\label{eqn:ratio}
\end{equation}

\begin{figure}[htbp] 
   \centering
   \includegraphics[width=7cm]{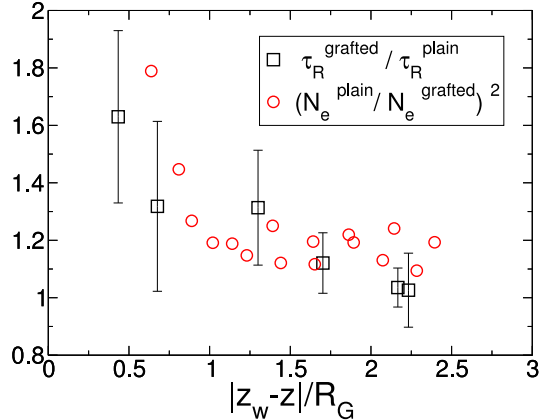}
   \caption{Comparison between the ratio of the local relaxation times
(left term in eq. \ref{eqn:ratio}) and
the inverse square ratio of the local entanglement lengths
(right term in eq. \ref{eqn:ratio}) of
the $N=100$ systems with and without grafted chains.}
   \label{fig:ratio}
\end{figure}

This relation provides a way to predict dynamics from a static
equilibrium quantity in the melt. It is reasonably well  verified
in the interfacial layer of our $N=100$ system, even if the melt
is only weakly entangled at this chain length (see fig.
\ref{fig:ratio}).
This argument can be pushed further to predict the local viscosity
in the interfacial layer in the reptation regime:
\begin{equation}
\frac{\eta^{grafted}}{\eta^{plain}} = \left(\frac{N_e^{plain}}{N_e^{grafted}}\right)^3
\label{eqn:ratio-eta}
\end{equation}
This prediction is no longer valid for chain lengths as short as $N=100$ as
we have only one entangled mode (too few entanglements per chain),
but should be verified for strongly
entangled melts where the mechanical behavior is entirely dictated
by the entanglements.

\section{Discussion and conclusions}

We presented here a study of  the dynamic and static behavior for
a polymer melt at the interface with a repulsive wall and a
repulsive wall subject to surface treatment creating about $20\%$
of chemisorbing sites on the surface. We find that the
modifications in the dynamics due to the surface treatment depends
on the level of entanglement of the melt. In the case of
unentangled melts, the dynamics is accelerated compared to the
bulk  due to the repulsive surface. The grafted chains locally
slow down the dynamics in the surface plane compared to systems
without grafted chains, but the relaxation times are smaller than
in the bulk and there is a local decrease in the viscosity as in
the case of a plain repulsive wall. In the case of weakly
entangled melts, the presence of the repulsive surface induces a
decrease ($\sim 30-40\%$) in the entanglement density in a range
of about $R_G$ from the surface. The presence of the grafted
chains prevents this depletion of the entanglements and further
reduces the entanglement length in the interfacial layer, in part
due to the smaller chain dimensions. This leads to slower dynamics
in the interfacial layer and a local increase of the relaxation
times of the entangled modes near the surface. The behavior of the
relaxation times can be predicted by measuring the local
entanglement length, following the expression given by  reptation
theory.

Our results are obtained for chains that are only slightly above
the entanglement threshold.  We expect however that these
conclusions would hold  for more entangled  melts. For longer
chains the extension of the region where entanglement effects are
enhanced by the grafting will be larger. It is likely that the
motion of the chains perpendicular to the interface will be slowed
down as well within this region, as entanglements will also hinder
this type of motion.

We note that the explanation of reinforcement by local variation 
of the entanglement length we proposed here is also relevant in a 
wider range of systems. We considered a surface with a certain 
number of infinitely attractive sites, but similar effects can be
expected for a surface with some attractive sites or a globally 
attractive surface. Thus the slowdown mediated by the adsorbed, slowest 
chains plays an important role in the mechanical properties of
nanocomposite polymer based materials.

Finally we would like to thank Pr. Ralf Everaers for fruitful discussions
concerning the primitive path analysis and the R\'egion Rh\^one-Alpes for 
a grant of computer time on the CCRT-CEA calculator.


\begin{thebibliography}{99}


\bibitem{payne} Payne, A.R.
{\em J. Appl. Pol. Sci.}  \textbf{1962}, {\em 6}, 57.

\bibitem{oberdisse} For a recent review see: Oberdisse J.
{\em Soft Matter} {\bf 2006}, {\em 2},  29.


\bibitem{sternstein} Sternstein, S. S.; Zhu, A.-J.; {\em Macromolecules} {\bf 2002}, {\em 35}, 7262.

\bibitem{gautier} Dalmas F, Chazeau L, Gauthier C, Cavaille JY, Dendievel
R.; {\em Polymer} {\bf 2006}, {\em 47 (8)}, 2802-2812.

\bibitem{lequeux}
 Berriot J, Montes H, Lequeux F, Long D, Sotta P
 {\em Europhys. Lett.} \textbf{2003}, {\em 64}, 50.


\bibitem{bashnagel-review} Baschnagel, J.; Varnik, F.; {\em J. Phys.: Condens. Matter} {\bf 2005},
 {\em 17}, 851.



\bibitem{Itagaki} Itagaki, H.; Nishimura, Y.; Sagisaka, E.; Grohens, Y.; {\em Langmuir} {\bf 2006}, {\em 22}, 742.

\bibitem{brown1} Brown, H. R.; Russell, T. P.; {\em Macromolecules} {\bf 1996}, {\em 29}, 798.

\bibitem{brown2} Oslanec, R.; Brown, H. R.; {\em Macromolecules} {\bf 2003}, {\em 36}, 5839.

\bibitem{binder-Rg} Wang, J.-S.; Binder, K.; {\em J. Phys. France} {\bf 1991}, {\em 1}, 1583.

\bibitem{kumar-Rg} Kumar, S. K.; Vacatello, M.; Yoon, D. Y.; {\em Macromolecules} {\bf 1990}, {\em 23}, 2189.

\bibitem{everaers} Everaers, R.; Sukumaran, S. K.; Grest, G. S.; Svaneborg, C.; Sivasubremanian, A.; Kremer, K.; {\em Science} {\bf 2004}, {\em 203}, 823.

\bibitem{sukumaran} Sukumaran, S. K.; Grest, G. S.; Kremer, K.; Everaers, R.; {\em J. Poly. Sci} {\bf 2005}, {\em 43}, 917.

\bibitem{vladkov-barrat} Vladkov, M.; Barrat, J.-L.; {\em Macromol. Th. Simu.} {\bf 2006}, {\em 15}, 252.

\bibitem{KremerGrest} K. Kremer, G. S. Grest,{\em J. Chem. Phys.} {\bf 1990}, {\em 92}, 5057.

\bibitem{Flory} Flory, P. J.; {\em Statistical Mechanics of Chain Molecules}, Interscience, Ney York, 1969.

\bibitem{mapping} Nath, S. K.; Frischknecht, A. L.; Curro, J. G.; McCoy, J. D.; {\em Macromolecules} {\bf 2005}, {\em 38}, 8562.

\bibitem{kurtsmith} Smith, K. A.; Vladkov, M.; Barrat, J.-L.; {\em Macromolecules} {\bf 2005}, {\em 38}, 571.

\bibitem{huber} Huber G.; Vilgis, T. A.; {\em Eur. Phys. J. B} {\bf 1998}, {\em 3}, 217.












\bibitem{lammps} S. J. Plimpton, {\em J. Comp. Phys.} {\bf 1995}, {\em 117}, 1.
LAMMPS web site: www.cs.sandia.gov/$^{\sim}$sjplimp/lammps.html

\end{thebibliography}
\end{document}